\def\sm{SM}

\newcommand{\beq}{\begin{equation}} 
\newcommand{\eeq}{\end{equation}}
\newcommand{\beqa}{\begin{eqnarray}}
\newcommand{\eeqa}{\end{eqnarray}}

\def\lesssim{{~\raise.15em\hbox{$<$}\kern-.85em
          \lower.35em\hbox{$\sim$}~}}                                  
         
\def\sign{\mbox{{\rm sign}}}
\def\al{\alpha}
\def\be{\beta}

\def\prd#1{Phys.\ Rev.\ D{\bf #1}}

\documentstyle[12pt,axodraw]{article}
\begin{document}
\begin{titlepage}

\large
\centerline {\bf Beyond the standard model with $B$ physics}
\normalsize
 
\vskip 2.0cm
\centerline {Yuval Grossman\footnote{yuvalg@physics.technion.ac.il}}
\vskip 1.0cm
\centerline {\it
Department of Physics, Technion--Israel Institute of Technology} 
\centerline{\it Technion City, 32000 Haifa, Israel}
\vskip 4.0cm
 
\centerline {\bf Abstract}
\vskip 1.0cm
Extensions of the Standard Model may have significant effects
on $B$ physics observables. Two examples of
methods that may find such effects are reviewed:
Resolving discrete ambiguities in CP asymmetries and
detecting right handed currents in radiative $B$ decays.
\vfill
\end{titlepage}

\newpage

\section{Introduction}


The ultimate goal of the $B$ physics program is 
to find inconsistencies within the Standard Model (SM) \cite{review}, 
in particular, to find indications for new physics 
in the flavor and CP violation sectors.
Therefore, we should maximize our ability to detect such effects, 
should they occur. It is the purpose of this talk to describe
some ideas in this direction. These ideas emerge from a model independent
approach, namely, to look for 
inconsistencies within the SM in as many ways as possible.
In general, new physics can modify the
SM effective operators and create new ones that are absent in the SM.
In particular, it can add new CP violating phases and
significantly modify 
the flavor changing neutral current operators.

Some of the SM predictions, however, cannot be significantly modified by 
new physics. 
It is likely that new short distance physics 
does not significantly change the SM
prediction that decay amplitudes that involve the spectator quark,
for example, $B_d \to D_s K$, are small.
It is also unlikely for new physics to significantly modify amplitudes
that are large in the SM. For example, CKM unsuppressed tree level
decays, e.g. $b \to c \bar u d$, 
are likely to be dominated by the SM amplitude. 
Generically, we expect new amplitudes to be significant
only when they compete with loop or highly CKM suppressed SM amplitudes. 

Thus, in order for an observable to provide a useful probe of new physics
the following three conditions have to be satisfied:
\begin{itemize}
\item it should be practically measurable; 
\item it should have small SM uncertainties; 
\item it is likely to be sensitive to new physics.
\end{itemize}
Clearly, an impractical to measure observable
is not a good candidate for new physics searches. 
Similarly, an observable that within the
uncertainties of the SM can assume any value, is not sensitive to new physics
effects. 
For example, the measurement of $\sin 2 \alpha$, which is 
very important within the
SM framework, is not very sensitive to new physics due to the 
large uncertainties in its value within the SM.
Finally, we like to look for observables where new physics effects are likely
to occurs. Ideally, we like to test all the SM predictions.
In practice, however, we cannot do it, and in many cases we have to 
assume the SM. For example, it is very unlikely that new physics will change 
the SM prediction that the CP asymmetries in $B \to \psi K_S$
and $B \to \psi K_L$ are opposite in sign. Therefore,
the two data samples are combined to enlarge the statistics.

In this talk we give two examples,  resolving discrete ambiguities
and searching right handed currents in $B \to s \gamma$ decays, 
that roughly satisfy the above conditions and thus can be used to search 
for new physics.

\section{Resolving discrete ambiguities}

If we assume CKM unitarity there are two independent angles
in the ``unitarity triangle'', both of which are related to the 
underlying non-zero phases of CKM matrix elements. 
We take them to be $\alpha$ and $\beta$ and we 
use the definition $\gamma = \pi - \beta - \alpha$.
In $B$ factory experiments we seek to measure quantities that, in the
absence of physics from beyond the SM, are simply related to
these angles.  Ignoring for the moment the effects of subleading
amplitudes, time dependence CP violating
asymmetries are proportional to $\sin2\phi$ where $\phi$ is
one of the angles of the unitarity triangle. 
In particular, the first two CP
asymmetries to be measured are likely to be in $B \to \psi K_S$ which
measures $\sin2\beta$, and in $B \to \pi^+\pi^-$ which measures
$\sin2\alpha$.  

However, a measurement of $\sin2\phi$ can only determine the
angle $\phi$ up to a four fold ambiguity: $\{\phi, \pi/2-\phi, \pi + \phi,
3 \pi/2 -\phi\}$ with the  
angles defined by convention to lie between $0$
and $2\pi$.  Thus, with two independent angles, there can be a priori a
total 16 fold ambiguity in their values as determined from CP asymmetry
measurements.  These ambiguities can limit our ability to test the
consistency between the measured value of these angles and the range
allowed by  other measurements interpreted in terms of the SM.
Within the SM, the present data on the CKM matrix
imply that $2\beta$
is in the first quadrant ($0 < \beta < \pi/4$), that $0 <\alpha< \pi$, and
that there is a correlation between the values of $\al$ and $\be$ \cite{gn1}.
Thus, among the 16 possible solutions at most two, 
and probably only one, will
be found to be consistent with \sm\ results.
Namely, the SM can predict the values of $\beta$ and almost always
of $\alpha$
once $\sin 2 \beta$ and $\sin 2 \alpha$ are measured.

In the presence of physics beyond the \sm\ the values of the
``would be'' $\alpha$ and $\beta$ extracted from asymmetry measurements may
not fall within their \sm\ allowed range. Such new physics cannot be
detected if the values of the asymmetry angles happen to be related via the
ambiguities to values that do overlap the SM range. Clearly, the
fewer ambiguous pairings that remain, the better our chance of recognizing
new physics should it occur. 

In order to resolve these ambiguities 
in addition to the values of $\sin 2\phi$,  
only the signs of $\cos 2 \phi$ and $\sin \phi$ for both
$\phi=\alpha$ and $\phi=\beta$ need to be determined.  These four signs
resolve the ambiguities completely:
$\sign(\cos 2 \phi)$ is used to resolve the $\phi \to \pi/2 - \phi$
ambiguity and
$\sign(\sin \phi)$ is used to resolve the $\phi \to \pi + \phi$
ambiguity.  
Several measurements which can determine $\sign(\cos 2 \phi)$ 
have been proposed \cite{GQ}. 
Uncertainties in calculation of hadronic effects do not
affect the interpretations of these measurements,
although they do depend on the known value of hadronic
quantities such as the width and the mass of the $\rho$.
The determination of
$\sign(\sin\phi)$, however, cannot be achieved
without some theoretical input on hadronic physics.  Quantities that are
independent of hadronic effects  always appear as the ratio of a product of
CKM matrix elements to the complex conjugate of the same product. Such pure
phases are thus always twice the difference of phases of the CKM elements.
Any observable that directly involves a weak phase difference of two CKM
elements, $\phi$, (rather than $2 \phi$) also involves hadronic quantities
such as the ratio of magnitudes of matrix elements and 
the difference of their strong
phases. Thus, in order to determine the sign of $\sin\alpha$ or $\sin\beta$
some knowledge about hadronic physics is required.

Below we give two examples of ideas that can be used to resolve discrete
ambiguities.

\subsection{Cascade mixing and the sign of $\cos 2 \beta$}

In the first example we explain how to determine
$\sign(\cos 2 \beta)$ based
on cascade mixing \cite{cascade,kayser}.
The standard CP asymmetry in $B \to \psi K_S$
is sensitive to $\sin 2 \beta$. The reason is that the asymmetry
is generated by an interference between the
$B_H \to \psi K_S$ and $B_L \to \psi K_S$ decay amplitudes. 
[$B_H$ ($B_L$) is the heavy (light) mass eigenstate.]
One of the amplitudes is proportional
to $\sin \beta$ and the other to $\cos \beta$. The interference
is proportional to their product, namely to $\sin 2 \beta$.
Sensitivity to $\cos 2 \beta$ arise when two amplitudes 
that are proportional to  $\sin \beta$ (or  $\cos \beta$) interfere. 
[Recall $\cos^2 \beta = 1-\sin^2\beta = (1 + \cos 2 \beta)/2$.]
The idea of \cite{kayser} is to use
interferences between the
$B \to \psi K_L$ and $B \to \psi K_S$ decay amplitudes.
Then, there are a total of four amplitudes, two that are proportional to  
$\sin \beta$ and two that are proportional to  $\cos \beta$.
When all the four amplitude interfere, there are terms that are proportional
to $\cos^2\beta$ and $\sin^2\beta$.

The problem is that we cannot use the full $B \to \psi K$ sample. 
We cannot use events where the kaon decay into a final state that
identifies its mass. For example, we identify  
a kaon that decays at time $t_K \gg 1/\Gamma_S$ as a $K_L$. 
[$\Gamma_S$ is the $K_S$ width.]
For such decays there are no $K_L$ -- $K_S$ interferences. 
The effects of $K_L$ -- $K_S$
interferences can be seen only in events where the
kaon decays into a final state that is common to
both $K_L$ and $K_S$. The best candidates are semileptonic decays. 
As long as $t_K \lesssim 1/\Gamma_S$
there is significant
interference between the $K_L$ and the $K_S$ components.
The time dependence of the decay chain is given by \cite{kayser}
\beqa
&&\Gamma(B(\bar B) \to \psi (\pi^- \ell^+ \nu))  \propto 
e^{-\Gamma_B t_B} \Big\{
e^{-\Gamma_S t_K} \left[1 \mp \sin 2\beta \sin(\Delta m_B t_B) \right] +  
\nonumber\\&&
e^{-\Gamma_L t_K} \left[1 \pm\sin 2\beta \sin(\Delta m_B t_B) \right] 
\pm 2e^{-{1 \over 2}(\Gamma_L+\Gamma_S) t_K} \nonumber \\&&
\big[\cos(\Delta m_B t_B)\cos(\Delta m_K t_k) + 
\cos 2\beta \sin(\Delta m_B t_B)\sin(\Delta m_K t_k) \big] \Big\} \,,
\eeqa
where $t_B$ ($t_K$) is the time where the $B$ ($K$) decay and 
$\Gamma_S$ ($\Gamma_L$) is the $K_S$ ($K_L$) width.
Note the term that is proportional to $\cos 2\beta$.
It is this term that enables us to resolve the discrete ambiguity.

\subsection{$B \to \psi K_S$ vs $B \to D^+D^-$}

In the second example we explain how to determined 
$\sign(\sin \beta)$ by comparing the CP asymmetries in 
$B \to \psi K_S$ and $B \to D^+D^-$.
We can write any decay amplitude as a sum of two amplitudes \cite{gq1}:
a tree-dominated amplitude
$A_T$, and a penguin-only amplitude, $A_P$, 
with a different weak phase. Then, we define
$r \equiv A_P / A_T$.
In the case of the angle $\beta$ we have one class of measurements, from 
$b \rightarrow c \bar c s$  processes such as $B \to \psi K_S$, that have
very small $r$.
For these channels the
CP asymmetry measurement determines $\beta$ up
to the usual four-fold ambiguity 
\beq \label{aPSIKS}
a_{\psi K_S} = -\sin 2 \beta.
\eeq
The other class of measurements is from $b \to c \bar c d$ decays such as
$B \to D^+ D^-$.
In this case we expect $r$ to be significant.
To leading order in $r$ we get
\beq \label{aDD}
a_{D^+D^-} =
\sin 2\beta - 2r\cos 2\beta \sin\beta\cos\delta.
\eeq
where $\delta$ is the strong phase difference between $A_T$ and $A_P$.
Comparing Eqs. (\ref{aDD}) and (\ref{aPSIKS}) we find
\beq \label{signfi}
a_{\psi K_S} + a_{D^+D^-} =
-2 r\cos\delta (\cos 2\beta \sin\beta).
\eeq
It is clear from this expression that we can fix the sign of
$\sin\beta$ only if we know the sign of $\cos 2 \beta$ and, in
addition, the sign of $r\cos\delta$. We assume the first of these
can be determined experimentally by 
one of the methods mentioned before.
Currently, there is no reliable way to determine the sign of 
$r\cos\delta$. 
In order to proceed, we must assume a model for hadronic calculation. 
Assuming factorization and that the top 
penguin is dominant, we get $r < 0$. 
Within the factorization approximation the relevant strong phases (almost)
vanish, so that $\delta \simeq 0$, and hence the sign of 
$r\cos\delta$ is given by the sign of $r$. 
Assuming $r\cos\delta<0$
as given by the factorization calculation we get
\beq \label{signbeta}
\sign(a_{\psi K_S} + a_{D^+D^-}) = 
\sign (\cos 2\beta \sin\beta).
\eeq
Note, in particular, that the \sm\ predicts
$\cos 2\beta \sin\beta > 0$, and therefore also that the asymmetry in
$D^+D^-$ is smaller in magnitude than the asymmetry in $\psi K_S$ 
(and opposite in sign). 

\section{Right handed currents in $b \to s \gamma$ decays}

The measurements of both exclusive and inclusive 
$b\to s\gamma$ decay rates are in good agreement with the SM predictions.
This still leaves open the possibility that new physics could 
be present, but it manifests itself only in the details of the 
decay process. For example, the SM predicts
that the photons emitted in $b\to s\gamma$ decays are
predominantly left-handed. This property does not hold true in
extensions of the SM such as the left-right symmetric models,
and therefore, can be used to signal the presence of new physics.
Unfortunately, in $B\to K^*\gamma$ 
decay all photon polarization information 
contained in the final hadron is lost. 
Therefore, other ways have to be used in order to extract this information.
Here be summarize four different ways that have been proposed. 

Measuring mixing-induced CP asymmetries in the inclusive $b\to s\gamma$ decay
was proposed as  an indirect method for probing photon polarization 
effects in \cite{Atwood}.
Since both $B$ and $\bar B$ must decay to a common final state,
the resulting asymmetry measures the interference of right- and
left-handed photon amplitudes. As the SM predicts a very small
right-handed admixture of photons in $b\to q\gamma$ decays, a large
mixing-induced CP asymmetry is a signal of new physics.
In \cite{MaRe} the $\Lambda_b \to \Lambda \gamma$
decay was studied. Since the decaying particle is
a fermion, the photon polarization information can be
indirectly extracted using the
$\Lambda$ polarization.

Another way of measuring the photon
polarization in the exclusive $B\to V\gamma$ decay makes use 
of the conversion electron pairs formed by the primary photon.
Electron--positron pairs from photons that were
produced in the inner part of the detector can be traced and their
production plane can be reconstructed with high accuracy. 
The angular distribution in the relative angle of the
$K^*\to K\pi$ decay plane and that of the conversion pair can be used
to determine the helicity amplitudes in the $B\to V\gamma$ decay 
\cite{gp}.
Actually, the photon does not have to be on shell.
The photon can be off shell and we can use 
the corresponding direct decay
$B\to Ve^+ e^-$ in the region where 
the dilepton invariant mass is close to the threshold since there 
photon exchange dominates the decay \cite{Mel}.

\section{Conclusions}
Many $B$ physics measurements are
aimed to study SM parameters. These measurements also be used to look for
physics beyond the SM since by overconstraning the unitarity triangle we
may be lucky to find indications for new physics. 
The measurements we mentioned in this talk, as well as many others, belong
to a different class. They are aimed to look for new physics
effects. They try to confirm SM predictions:
that $\cos 2 \beta$ is positive, that the
photon in $b \to s \gamma$ is left handed, that the semileptonic
CP asymmetry in $B_s$ decays is very small, and many other.
Since a major goal of the $B$ physics program is to look for
new physics it is important to find and carry out 
this kind of measurements.

\section*{Acknowledgments}
I thank Dan Pirjol and Helen Quinn for pleasant collaborations on these 
subjests and Yossi Nir for comments on the manuscript.


\end{document}